\documentclass{article}
\usepackage{spconf,amsmath,graphicx}
\usepackage{amsfonts}
\usepackage{bm}
\usepackage{subfigure}
\usepackage{float}
\usepackage{tikz,pgfplots}
\usepackage{bm}
\usepackage{cases}
\usepackage{cleveref}
\usepackage{algorithm}
\usepackage{algorithmic}
\usepackage{multirow}
\usepackage{cite,bbm,booktabs,url}

\usepackage{shortcuts_OPT}

\newtheorem{Lemma}{Lemma}

\newtheorem{Corollary}{Corollary}

\title{Estimating Centrality Blindly from Low-pass Filtered Graph Signals}
\name{Yiran He, Hoi-To Wai}
\address{Department of SEEM, The Chinese University of Hong Kong, Shatin, Hong Kong SAR of China}
\begin{document}
\ninept
\maketitle
\begin{abstract}
This paper considers \emph{blind} methods for centrality estimation from graph signals. We model graph signals as the outcome of an unknown low-pass graph filter excited with influences governed by a sparse sub-graph. This model is compatible with a number of data generation process on graphs, including stock data and opinion dynamics.
Based on the said graph signal model, we first prove that the folklore heuristics based on PCA of data covariance matrix may fail when the graph filter is not sufficiently low-pass. To remedy, we propose a robust blind centrality estimation method which substantially improves the centrality estimation performance. Numerical results on synthetic and real data   support our findings.
\end{abstract}
\section{Introduction}
\label{sec:intro}

%

Graph learning is the problem of recovering pairwise relationships (edges) between nodes, based on observations of the nodes' behaviors. This is a problem of growing interest in many disciplines such as social science, biology and finance. Without knowing the exact correspondence or model between the ``graph'' of interest and these behaviors, the graph learning problem is \emph{ill-posed} in general. That said, using the intuition that connected nodes behave similarly, existing heuristics deploy the covariance or correlation matrices of the nodes' behavior as a proxy to the graph adjacency matrix \cite{kolaczyk2014statistical}. Though such a proxy matrix is easy to compute, it is not clear if such heuristics can produce valid results.
To remedy, previous works have considered models such as Gaussian Markov Random Field \cite{friedman2008sparse}, or linear/non-linear dynamical systems \cite{timme2007revealing,wai2016active,ravazzi2017learning,wai2019joint,giannakis2018topology,matta2018consistent}. However, these methods often impose strong restrictions on the data model, and the learning methods involve significant computation efforts.

The emerging field of graph signal processing (GSP) \cite{sandryhaila2013discrete,shuman2013emerging} works with \emph{graph signals}   supported on the node set of graph. GSP has been applied to a number of tasks for processing graph signals, including sampling and interpolation \cite{marques2015sampling,anis2016efficient}, filtering \cite{sandryhaila2014big}, etc.. 
On the other hand, studying the \emph{inverse problems} of GSP has inspired new models for graph learning. We treat the observations as \emph{filtered graph signals} and learn the graph topology as a blind deconvolution problem. Several inference methods have been proposed, e.g., based on smoothness \cite{dong2016learning}, spectral template \cite{segarra2017network}, structures of topology \cite{egilmez2017graph} etc., see the overview in \cite{dong2019learning,mateos2019connecting}. 
Meanwhile, for practitioners of graph learning, obtaining the graph topology is only the \emph{first step}, and  oftentimes the end goal is to deduce interpretable features from the graph topology such as node centrality, communities. For instance, in the context of social network, we study the relative importance of individuals via evaluating the node centrality. Analyzing these feature as a \emph{second step} can be undesirable because of error propagation and the additional computation complexity.

To address the above issues, 
we derive a \emph{blind} method for estimating graph features directly from a set of filtered graph signals. Note a number of recent works \cite{wai2018blind,hoffmann2018community,schaub2019blind,yuan2019learning} have worked on the related problem of blind \emph{community detection}.
In contrary, this paper proposes to estimate \emph{node centrality} in a similarly blind manner.
Specificatlly this paper studies a model where the graph signals are generated from exciting a low pass graph filter \cite{wai2018blind} via a sparse influence matrix. Our model is motivated by a diffusion process in stock market. In this model, the stock prices are affected by a small number of types of trending news and each news affects only a few stocks. These news act as stimuli to an unknown low-pass graph filter and we observe the output as the daily return. To this end, our method finds the most influential stocks through estimating the centrality of stocks in a latent inter-stock influence network.
 
Our contributions are three-fold. First, we show that the folklore heuristics of principal component analysis (PCA) on graph signal covariance matrix may produce inaccurate centrality estimation when the graph filter is not sufficiently \emph{low-pass}. Second, we show that when the PCA-based estimation fails, a robust method can be applied to improve the estimation quality. Lastly, we illustrate the effectiveness of the proposed method on synthetic and stocks data. For the latter, we utilize scores from Google Trend to approximate the activity levels of trending news.
\vspace{.1cm}

\noindent \textbf{Notations.} Boldfaced (capital) letters are used to define vectors (matrices), and we use $(\cdot)^\top$ to denote matrix/vector transpose. $\Re, \ZZ$ is the set of real and integer numbers. For a symmetric matrix ${\bm A}$, ${\sf TopEV}({\bm A})$ denotes its eigenvector with the largest eigenvalue.\vspace{-.2cm}

\section{Graph Signal Model}\vspace{-.2cm}
\label{sec:gsmodel}
In this section, we first discuss a graph signal model with excitations modeled through a sparse influence matrix, then describe a case study on modeling stocks data. 
Consider an $n$ nodes undirected graph $G = (V,E)$ with the node set $ V := \{1, \dots, n\}$ and edge set $ E \subseteq V \times V $. From the undirected graph \emph{G}, we define a symmetric adjacency matrix ${\bm A} \in \Re^{n \times n}$ such that $ A_{ij} = A_{ji} =1$ if and only if $(i,j) \in E $; otherwise, $A_{ij} = 0$. The symmetric matrix ${\bm A}$ admits an eigenvalue decomposition (EVD) as ${\bm A} = {\bm V} \bm{\Lambda} {\bm V}^\top$, where ${\bm V}$ is an orthogonal matrix and $\bm{\Lambda} = {\rm Diag}( \bm{\lambda} )$ is a diagonal matrix with $\lambda_1 \geq \lambda_2 \geq \cdots \geq \lambda_n$.  
We study the eigen-centrality defined as:\vspace{-.2cm}
\begin{equation} \label{eq:eigcen}
{\bm c}_{\sf eig} = {\sf TopEV}({\bm A}) = {\bm v}_1.\vspace{-.2cm}
\end{equation}
If $G$ is a connected undirected graph, then by the Perron-Frobenius theorem \cite{horn2012matrix}, ${\bm c}_{\sf eig}$ must be a positive vector. Hence, the vector ${\bm c}_{\sf eig}$ induces an ordering of the nodes $V$. This ordering is regarded as a measure of the \emph{centrality}, and it is related to a number of common measures such as PageRank, Katz centrality, see \cite{vigna2016spectral}. \vspace{.1cm}

\noindent \textbf{Graph Signals and Filter.}
A graph signal is defined as a function supported on $V$, \ie $x: V \rightarrow \RR$, which can be succinctly described as an $n$-dimensional vector.
Consider the two graph signals: ${\bm x} = [x_1, \dots, x_n]^\top \in \Re^n$ and ${\bm y} = [y_1, \dots, y_n]^\top \in \Re^n$, where $x_i$ and $y_i$ represent the signal values of node $i$ at the initial conditions and in the filtered conditions, respectively. The \emph{graph filter} describes the process of network/graph $G$ that maps ${\bm x}$ to ${\bm y}$. Using the adjacency matrix ${\bm A}$ as a graph shift operator, the (linear)  \emph{graph filter} ${\cal H}( {\bm A})$ can be expressed as a matrix polynomial of  order $T$ as:
\begin{equation} \label{eq:graphfil}
{\cal H}({\bm A}) = \sum_{t=0}^{T-1} h_t {\bm A}^t = {\bm V} \left(\sum_{t=0}^{T-1}h_t \bm{\Lambda}^t \right) {\bm V}^\top \in \Re^{n\times n}.
\end{equation}
We have defined the generating polynomial as $h(\lambda):= \sum_{t=0}^{T-1}h_t\lambda^t$ from the filter coefficients $\{ h_t \}_{t=0}^{T-1}$ and used the EVD of ${\bm A}$ to derive the last equality.
Given an input graph signal ${\bm x} \in \Re^n$, the output, ${\bm y} \in \Re^n$, of the graph filter ${\cal H}({\bm A})$ is a linear combination of the local diffused versions of ${\bm x}$. In other words, the output ${\bm y}$ is expressed as:\vspace{-.1cm}
\begin{equation} \textstyle \label{eq:y=Hx}
{\bm y} =  {\cal H}( {\bm A} ){\bm x} = \sum_{t=0}^{T-1} h_t {\bm A}^t {\bm x},\vspace{-.1cm}
\end{equation}
where $ {\bm A}^t {\bm x}$ represents a $t$-hop diffusion of the input signal ${\bm x}$ on $G$.
 
Throughout this paper, we assume the following strict inequality on the spectrum of ${\cal H}({\bm A})$:
\beq \textstyle \label{eq:lowpass}
|h(\lambda_1)| > \max_{j=2,...,n} |h(\lambda_j)|.
\eeq 
This condition is also known as the \emph{$1$-low pass} in the author's prior work \cite{wai2018blind} which implies that the vector ${\bm c}_{\sf eig} = {\bm v}_1$ remains the top eigenvector of the graph filter ${\cal H}( {\bm A})$.
This condition is common for processes on graphs in practice. For example, the $T_{\sf dif}$-step diffusion processes ${\cal H}({\bm A})  = ( \alpha_0 {\bm I} + \alpha_1 {\bm A})^{T_{\sf dif}}$ 
and the infinite impulse response (IIR) filter ${\cal H}({\bm A})=({\bm I} - c{\bm A})^{-1}$ which models equilibriums of quadratic games on networks or DeGroot opinion dynamics with stubborn agents.\vspace{.1cm}

\noindent \textbf{Sparse Influence Model.}
Without loss of generality, the input graph signal ${\bm x}$ to the graph filter ${\cal H}({\bm A})$ can be written as the product between a matrix ${\bm B} \in \Re^{n\times k}$ and a vector ${\bm z} \in \Re^k$:\vspace{-.1cm}
\begin{equation}\label{eq:x=Bz} \textstyle
 {\bm x} = {\bm {Bz}} = \sum_{j=1}^k {\bm b}_j z_j.\vspace{-.1cm}
\end{equation} 
In other words, the matrix ${\bm B}$ defines the range space of the input graph signal ${\bm x}$. Specifically, the column vector ${\bm b}_j$ can be viewed as the influence profile from an external source $z_j$, to impact the graph, the input graph signal is a combined effect of these influences. 
We argue that ${\bm B}$ is a \emph{sparse matrix} in some applications -- when the influence from external sources are \emph{localized}.  
\vspace{-.1cm}


\subsection{Case Study: Inter-stock Influence Network} \label{sec:stock}\vspace{-.1cm}
We consider a dataset of daily return (\ie the ratio between closing and opening price) for $n$ stocks recorded on $m$ days, denoted as $\{ {\bm y}_i \}_{i=1}^m$. As a hypothesis, for each day $i$, the daily return ${\bm y}_i$ is a filtered graph signal following \eqref{eq:y=Hx}. Here, the graph filter ${\cal H}({\bm A})$ is defined on the inter-stock influence network, which encodes the strengths of influences among stocks reflected in their performance. While neither ${\cal H}( {\bm A} )$ nor the coefficients $\{ h_t \}_{t=0}^{T-1}$ is known, we expect it to satisfy \eqref{eq:lowpass} as the process is akin to diffusion. On the other hand, the input graph signal ${\bm x}_i$ follows the sparse influence model.

The stock market is affected by external stimuli/sources {such as policy changes}, it is hence reasonable to model the input graph signal ${\bm x}_i$ as an outcome from mapping the \emph{state-of-the-world} to an impact on the stock market.
Specifically, we have ${\bm x}_i = {\bm B} {\bm z}_i$ to model the direct \emph{impact} and each element of ${\bm z}_i$ denotes the state-of-the-world from a specific type, e.g., `oil crisis', `trade war', etc.. The latter can be estimated from sources which measures the popularity of a key word, e.g., via Google Trend.
Here, the matrix ${\bm B}$ is sparse as the state-of-the-world generates only localized effects on the stocks, e.g., a technology company is less likely to be affected by the `oil crisis', leading to a sparse vector for the corresponding `${\bm b}_j$' [cf.~\eqref{eq:x=Bz}]. 

We are interested in learning about the inter-stock influence network ${\bm A}$, specifically the centrality ${\bm c}_{\sf eig}$ which reveals the relative \emph{intrinsic influence powers} of stocks in the market. However, learning the complete graph topology is difficult, if not impossible, since the underlying filter coefficients $\{ h_t \}_{t=0}^{T-1}$ and influence structure ${\bm B}$ are clearly unknown. 
This motivates the current paper to propose a \emph{blind} method {only} for centrality estimation, as detailed next.\vspace{-.1cm}



  
\section{Blind Centrality Estimation}\vspace{-.1cm}
\label{sec:CenEst}

We study the problem of estimating eigen-centrality of nodes when the graph adjacency matrix ${\bm A}$, filter coefficients $\{ h_t \}_{t=0}^{T-1}$, and influence model ${\bm B}$, are unknown. 
Instead, our estimation procedure is based on indirect observations of the graph with  the filtered signals $\{ {\bm y}_i \}_{i=1}^m$ and the corresponding latent parameter vectors $\{ {\bm z}_i \}_{i=1}^m$, where each pair $( {\bm y}_i, {\bm z}_i )$ satisfies \eqref{eq:y=Hx}, \eqref{eq:x=Bz}.\vspace{.1cm}

\noindent \textbf{PCA-based Estimation}.
The covariance/correlation matrix of graph data is commonly recognized as a proxy of  graph adjacency matrix \cite{kolaczyk2014statistical}. Due to this interpretation, a natural idea is to apply principal component analysis (PCA) on the covariance of filtered graph signals and use the principal eigenvector as an estimate for ${\bm c}_{\sf eig}$. Here, we analyze the performance of such procedure using GSP model.  

We assume that ${\bm z}_i$ in \eqref{eq:x=Bz} is zero-mean and satisfies $\EE_i [ {\bm z}_i {\bm z}_i^\top ] = {\bm I}$. As ${\bm x}_i = {\bm B} {\bm z}_i$, the covariance of ${\bm y}_i$ is \vspace{-.1cm}
\beq
{\bm C}_y = \EE_i [ {\bm y}_i {\bm y}_i^\top ] = {\cal H}( {\bm A} ) {\bm B} {\bm B}^\top {\cal H}( {\bm A} )\vspace{-.1cm}
\eeq
Define the generating polynomial $h(\lambda) \eqdef \sum_{t=0}^{T-1}h_t{\lambda}^t$ and using \eqref{eq:graphfil},  we observe that  \vspace{-.1cm}
\beq
{\bm C}_y = {\bm V} h( \bm{\Lambda} ) {\bm V}^\top {\bm B} {\bm B}^\top {\bm V} h( \bm{\Lambda} ) {\bm V}^\top. \vspace{-.1cm}
\eeq
When ${\bm B} = {\bm I}$ and \eqref{eq:lowpass} holds, the top eigenvector of ${\bm C}_y$ will be given by the desired ${\bm c}_{\sf eig}$. In this case, centrality estimation can be solved by estimating $\widehat{\bm v}_1 \eqdef {\sf TopEV}({\bm C}_y)$.
However, in general we have ${\bm B} \neq {\bm I}$. 
Let $\widehat{\bm v}_1 \eqdef {\sf TopEV}( {\bm C}_y)$, we observe the following:
\begin{Lemma} \label{lem:perfbd}
Suppose that {\sf (i)} ${\bm v}_1^\top {\bm B} {\bm q}_1 \neq 0$, where ${\bm q}_1$ is the top right singular vector of ${\cal H}( {\bm A} ) {\bm B}$, and {\sf (ii)} $h(\lambda_1) > \max_{j=2,...,n} h(\lambda_j)$.
Then, it holds that
\beq \label{eq:perf_bd}
\begin{split}
& \min\{ \| {\bm c}_{\sf eig} -  \widehat{\bm v}_1 \|_2, \| {\bm c}_{\sf eig} + \widehat{\bm v}_1 \|_2 \} \\
& \leq \sqrt{2} \cdot
\frac{ \max_{j=2,...,n} |h(\lambda_j)| }{|h(\lambda_1)|} \cdot \frac{ \| {\bm V}_{N-1}^\top {\bm B} {\bm q}_1 \|_2 }{ {\bm v}_1^\top {\bm B} {\bm q}_1 }.
\end{split}
\eeq
\end{Lemma}
The above lemma is adapted and simplified from \cite[Proposition 1]{wai2018blind}. In the lemma,  {\sf (i)} is a regulatory condition on the influence matrix ${\bm B}$. The condition {\sf (ii)} requires that the graph filter ${\cal H}({\bm A})$ should maintain ${\bm v}_1$ as its top eigenvector.
A bound for $\| \widehat{\bm v}_1 \pm {\bm c}_{\sf eig} \|_2$ is in \eqref{eq:perf_bd} which is controlled by the low-pass property of graph filter. The error is reduced when $\frac{ \max_{j=2,...,n} |h(\lambda_j)| }{ |h(\lambda_1)| } \ll 1$, \ie the graph filter is \emph{sufficiently low-pass}.
It is worthwhile to mention the special case of ${\bm B} = {\bm I}$, where ${\bm q}_1 = {\bm v}_1$ and $\| {\bm V}_{N-1}^\top {\bm B} {\bm q}_1 \|_2 = 0$, implying $\widehat{\bm v}_1 = {\bm c}_{\sf eig}$. Moreover, in practice, only a sample covariance matrix is used in lieu of ${\bm C}_y$ for centrality estimation.\vspace{.1cm}

\noindent \textbf{Proof.} Our proof consists in recognizing that Condition~1 to 3 in \cite[Proposition 1]{wai2018blind} are satisfied by our {\sf (i)}, {\sf (ii)}. In particular, adapting the result therein with $K=1$, we obtain:\vspace{-.1cm}
\beq \label{eq:prop_bd}
\| \widehat{\bm v}_1 \widehat{\bm v}_1^\top - {\bm c}_{\sf eig} {\bm c}_{\sf eig}^\top \|_2
= \sqrt{ \gamma^2 / (1+\gamma^2) } \leq \gamma,
\eeq
where 
\beq
\gamma \leq \frac{ \max_{j=2,...,n} |h(\lambda_j)| }{ |h(\lambda_1)| } \cdot \frac{ \| {\bm V}_{N-1}^\top {\bm B} {\bm q}_1 \|_2 }{ {\bm v}_1^\top {\bm B} {\bm q}_1 }.
\eeq
On the other hand, if one has $\widehat{\bm v}_1^\top {\bm c}_{\sf eig} \geq 0$, then
\beq
\begin{split}
\| \widehat{\bm v}_1 \widehat{\bm v}_1^\top - {\bm c}_{\sf eig} {\bm c}_{\sf eig}^\top \|_2 & 
\geq (1/\sqrt{2}) \| \widehat{\bm v}_1 \widehat{\bm v}_1^\top - {\bm c}_{\sf eig} {\bm c}_{\sf eig}^\top \|_{\rm F} \\
& \geq (1/\sqrt{2}) \| \widehat{\bm v}_1 - {\bm c}_{\sf eig} \|_2,
\end{split}
\eeq 
otherwise if $\widehat{\bm v}_1^\top {\bm c}_{\sf eig} < 0$, then
\beq
\| \widehat{\bm v}_1 \widehat{\bm v}_1^\top - {\bm c}_{\sf eig} {\bm c}_{\sf eig}^\top \|_2 \geq (1/\sqrt{2}) \| \widehat{\bm v}_1 + {\bm c}_{\sf eig} \|_2 ,
\eeq
where we have used that $\| \widehat{\bm v}_1 \|_2 =1, \|{\bm c}_{\sf eig} \|_2 = 1$ in both inequalities. Substituting this in \eqref{eq:prop_bd}
yields the desired result.\hfill \textbf{Q.E.D.}\vspace{.2cm}

Let us now examine the practicality of \emph{sufficiently low-pass} condition, \ie $\frac{ \max_{j=2,...,n} |h(\lambda_j)| }{ |h(\lambda_1)| } \ll 1$, by studying specific graph filters. We are interested in the infinite impulse response (IIR) filter:\vspace{-.1cm}
\beq \label{eq:iir}
{\cal H}( {\bm A} ) = ( {\bm I} - \alpha {\bm A} )^{-1} = {\bm I} + \alpha {\bm A} + \alpha^2 {\bm A}^2 + \cdots,\vspace{-.1cm}
\eeq
where $\alpha >0$ a small constant satisfying $\alpha < \lambda_1^{-1}$. The generating polynomial is 
$h(\lambda) = (1-\alpha \lambda)^{-1} = 1 + \alpha \lambda + \cdots$, which is increasing, non-negative for $\lambda \in [ \lambda_n , \lambda_1]$. We have\vspace{-.1cm}
\beq
\frac{ \max_{j=2,...,n} |h(\lambda_j)| }{ |h(\lambda_1)| } = \frac{ h(\lambda_2) }{h( \lambda_1) } = \frac{ 1 - \alpha \lambda_1 }{ 1 - \alpha \lambda_2 },\vspace{-.1cm}
\eeq
which is close to one for small $\alpha$. Applying Lemma~\ref{lem:perfbd} indicates that the PCA-based centrality estimation may be inaccurate when the observations are generated from this class of graph filters.
\vspace{.1cm}

\noindent \textbf{Graph Filter Robust Estimation}. To design a centrality estimation method agnostic to the underlying graph filtering process, we consider the \emph{boosted} graph filter \cite{wai2018blind} defined as:\vspace{-.1cm}
\beq \label{eq:boost}
\widetilde{\cal H}_\rho ( {\bm A} ) \eqdef {\cal H}({\bm A}) - \rho {\bm I},\vspace{-.1cm}
\eeq
which has the generating polynomial as $\tilde{h}_\rho( \lambda ) \eqdef h(\lambda) - \rho$. If $h( \lambda )$ is positive on $\lambda \in [\lambda_n,\lambda_1]$, there exists $\rho>0$ with\vspace{-.1cm}
\beq
\frac{ \max_{j=2,...,n} | \tilde{h}_\rho (\lambda_j)| }{ | \tilde{h}_\rho (\lambda_1)| } \leq c_{\sf boost} \frac{ \max_{j=2,...,n} |h(\lambda_j)| }{ |h(\lambda_1)| },\vspace{-.1cm}
\eeq
where $c_{\sf boost} \in [0,1)$. It can be confirmed for the IIR filter described in \eqref{eq:iir} -- with $\rho=1$, we have $c_{\sf boost} = \frac{\lambda_2}{\lambda_1} < 1$. In the latter case, we have $c_{\sf boost} \ll 1$ if $\lambda_2 \ll \lambda_1$. This condition  holds when the graph has a few highly connected nodes. We also note that $\widetilde{\cal H}_\rho ( {\bm A} )$ is close to rank-one if $c_{\sf boost} \ll 1$.

With the improved low-pass property, it is desirable to work with the \emph{boosted filter} $\widetilde{\cal H}_\rho ( {\bm A} )$. 
To this end, we observe that\vspace{-.1cm}
\beq
{\bm Y} \equiv ( {\bm y}_1~\cdots~{\bm y}_m ) = {\cal H}({\bm A}) {\bm B} ( {\bm z}_1~\cdots~{\bm z}_m) \equiv {\cal H}({\bm A}) {\bm B} {\bm Z}.\vspace{-.1cm}
\eeq
With the pairs of filtered signals $\{ {\bm y}_i \}_{i=1}^m$ and latent vectors $\{ {\bm z}_i \}_{i=1}^m$ available,
it is straightforward to estimate $ {\cal H}({\bm A}) {\bm B}$ through:\vspace{-.1cm}
\beq \label{eq:hathb}
\widehat{{\bm H}{\bm B}} = {\bm Y} ({\bm Z}^\top {\bm Z})^{-\top} {\bm Z}^\top \approx {\cal H}({\bm A}){\bm B},\vspace{-.1cm}
\eeq 
where 
${\cal H}({\bm A}){\bm B}$ admits the low-rank+sparse decomposition as:\vspace{-.1cm}
\begin{equation}\label{eq:H=L+S}
{\cal H}({\bm A}){\bm B} = \widetilde{\cal H}( {\bm A} ) {\bm B} + \rho {\bm B} \equiv {\bm L}+{\bm S}.\vspace{-.1cm}
\end{equation}
In particular, let $\widetilde{\bm v}_1$ be the top \emph{left} singular vector of ${\bm L}$, we have
\begin{Corollary}\label{cor:perf}
Under the same conditions as Lemma~\ref{lem:perfbd}. It holds\vspace{-.1cm}
\beq \label{eq:perf_bd2}
\begin{split}
& \min\{ \| {\bm c}_{\sf eig} -  \widetilde{\bm v}_1 \|_2, \| {\bm c}_{\sf eig} + \widetilde{\bm v}_1 \|_2 \} \\
& \leq c_{\sf boost} \cdot  \sqrt{2} \cdot
\frac{ \max_{j=2,...,n} |h(\lambda_j)| }{|h(\lambda_1)|} \cdot \frac{ \| {\bm V}_{N-1}^\top {\bm B} {\bm q}_1 \|_2 }{ {\bm v}_1^\top {\bm B} {\bm q}_1 }.
\end{split}
\eeq
\end{Corollary}
The corollary can be obtained by repeating the analysis for Lemma~\ref{lem:perfbd} through replacing ${\cal H}( {\bm A} )$ with $\widetilde{\cal H}( {\bm A})$. Particularly, the corollary indicates that we can reliably estimate the node centralities from analyzing the singular vectors of ${\bm L}$, especially when $c_{\sf boost} \ll 1$.

Our final task is to compute the decomposition \eqref{eq:H=L+S} in order to retrieve ${\bm L}$. To this end, we observe ${\bm L}$ is close to rank-one if $c_{\sf boost} \ll 1$ and ${\bm S} \eqdef \rho {\bm B}$ is sparse under the considered influence model.
The matrix decomposition problem can be effectively tackled through the convex problem \cite{agarwal2012noisy}:\vspace{-.1cm}
\begin{equation}\label{eq:rpca}
\min_{ \hat{\bm L}, \hat{\bm S} }~\| \widehat{{\bm H}{\bm B}} - \hat{\bm L} - \hat{\bm S} \|_F^2 + \lambda_L \| \hat{\bm L} \|_\star + \lambda_S \| {\rm vec}(\hat{\bm S}) \|_1,\vspace{-.1cm}
\end{equation}
where $\| \cdot \|_1$, $\| \cdot \|_\star$ denotes the $\ell_1$, nuclear norm, and $\lambda_S, \lambda_L \geq 0$ are regularization parameters controlling the low-rankness and sparseness in the solution. 
An estimate of the centrality vector can then be computed from the top left singular vector of $\hat{\bm L}$. 
We conclude the section by discussing a few implementation issues.\vspace{-.1cm}

\algsetup{indent=1em}
\begin{algorithm}[tb]
	\caption{Graph Filter Robust Centrality Estimation.}\label{alg:boosted}
	\begin{algorithmic}[1]
		\STATE {\textbf{INPUT}}: graph signals $\{\bm y_i \}_{i=1}^m$ and latent parameters $\{ {\bm z}_i \}_{i=1}^m$.
		\STATE Compute/estimate the graph filter-influence matrix product as $\widehat{\bm{HB}}$ via the least square solution \eqref{eq:hathb}.
		\STATE Decompose $\widehat{\bm{HB}}$ via solving the convex problem \eqref{eq:rpca}.
		\STATE If quantized ${\bm B}$ is preferred, set $\overline{\bm L} = \widehat{\bm{HB}} - \widehat{\bm S}^{\sf thres}$; otherwise, set $\overline{\bm L} = \hat{\bm{L}}$
		\STATE {\textbf{OUTPUT}}: estimate of ${\bm c}_{\sf eig}$ as $\widetilde{\bm v}_1$, i.e., top left singular vector $\overline{\bm L}$.
	\end{algorithmic} 
\end{algorithm}

\begin{figure*}[t]
\centering
\resizebox{!}{.18\textheight}{\begin{tikzpicture}[scale=0.8]
\definecolor{color0}{rgb}{1,0.498039215686275,0.313725490196078}
\begin{axis}[
legend cell align={left},
legend style={draw=white!80.0!black},
tick align=outside,
tick pos=both,
title={\large {(a)} 10\% sparse ${\bm B}$},
x grid style={white!50!black},
xlabel={\large Dim.~of latent factor $k$},
xmajorgrids,
xmin=5.5, xmax=104.5,
xtick style={color=black},
y grid style={white!50!black},
ylabel={\large Error rate},
ymajorgrids,
ymin=0.00, ymax=0.75,
ytick style={color=black}, 
width=6cm,height=6cm
]
\addplot [semithick, black, mark=*, mark size=3, mark options={solid}]
table {%
10 0.6528
20 0.5516
30 0.4648
40 0.4148
50 0.3702
60 0.3402
70 0.3232
80 0.3008
90 0.2766
100 0.2708
};
\addlegendentry{PCA-based Est.}
\addplot [very thick, red, mark=triangle, mark size=3, mark options={solid}]
table {%
10 0.6528
20 0.4058
30 0.1364
40 0.0898
50 0.0728
60 0.0686
70 0.0712
80 0.0676
90 0.0636
100 0.0674
};
\addlegendentry{Robust Est.}
\addplot [very thick, blue, mark=square, mark size=3, mark options={solid}]
table {%
10 0.6528
20 0.4168
30 0.131
40 0.0654
50 0.0342
60 0.0204
70 0.017
80 0.01
90 0.00639999999999996
100 0.00680000000000003
};
\addlegendentry{Robust Est. (w/ Quant.)}
\end{axis}
\end{tikzpicture}}
\resizebox{!}{.18\textheight}{\begin{tikzpicture}[scale=0.6]
\definecolor{color0}{rgb}{1,0.498039215686275,0.313725490196078}
\begin{axis}[
legend cell align={left},
legend style={draw=white!80.0!black},
tick align=outside,
tick pos=both,
title={\large {(b) ${\cal U}(\{3,....,6\})$ non-zeros}},
x grid style={white!50!black},
xlabel={\large Dim.~of latent factor $k$},
xmajorgrids,
xmin=15.5, xmax=104.5,
xtick style={color=black},
y grid style={white!50!black},
ymajorgrids,
ymin=0.00, ymax=0.75,
ytick style={color=black}, 
width=6cm,height=6cm
]
\addplot [very thick, black, mark=*, mark size=3, mark options={solid}]
table {%
20 0.1496
30 0.3482
40 0.5202
50 0.628
60 0.6838
70 0.6962
80 0.7088
90 0.7238
100 0.7194
};
\addplot [very thick, red, mark=triangle, mark size=3, mark options={solid}]
table {%
20 0.148
30 0.0434
40 0.0308
50 0.0438
60 0.0556
70 0.072
80 0.0836
90 0.0996
100 0.116
};
\addplot [very thick, blue, mark=square, mark size=3, mark options={solid}]
table {%
20 0.1496
30 0.0392
40 0.0188
50 0.0128
60 0.0125999999999999
70 0.0125999999999999
80 0.00980000000000003
90 0.01
100 0.0108
};
\end{axis}
\end{tikzpicture}}
\resizebox{!}{.18\textheight}{\begin{tikzpicture}[scale=0.6]
\definecolor{color0}{rgb}{1,0.498039215686275,0.313725490196078}
\begin{axis}[
legend cell align={left},
legend style={draw=white!80.0!black},
tick align=outside,
tick pos=both,
title={\large {(c) ${\cal U}(\{0,....,\lceil 0.1k\rceil\})$ non-zeros}},
x grid style={white!50!black},
xlabel={\large Dim.~of latent factor $k$},
xmajorgrids,
xmin=5.5, xmax=104.5,
xtick style={color=black},
y grid style={white!50!black},
ymajorgrids,
ymin=0.00, ymax=0.75,
ytick style={color=black}, 
width=6cm,height=6cm
]
\addplot [very thick, black, mark=*, mark size=3, mark options={solid}]
table {%
10 0.6854
20 0.3392
30 0.263
40 0.241
50 0.2386
60 0.2298
70 0.2238
80 0.2234
90 0.2262
100 0.221
};
\addplot [very thick, red, mark=triangle, mark size=3, mark options={solid}]
table {%
10 0.6854
20 0.3082
30 0.137
40 0.111
50 0.1118
60 0.1238
70 0.13
80 0.1462
90 0.157
100 0.1618
};
\addplot [very thick, blue, mark=square, mark size=3, mark options={solid}]
table {%
10 0.6854
20 0.316
30 0.1322
40 0.0764
50 0.0564
60 0.0532
70 0.0412
80 0.04
90 0.0427999999999999
100 0.039
};
\end{axis}
\end{tikzpicture}}~~~
\begin{minipage}{.25\linewidth}
\begin{center}
{\scriptsize Setting (a)/(b)/(c), $k=20$} \\
\includegraphics[height=.08\textheight]{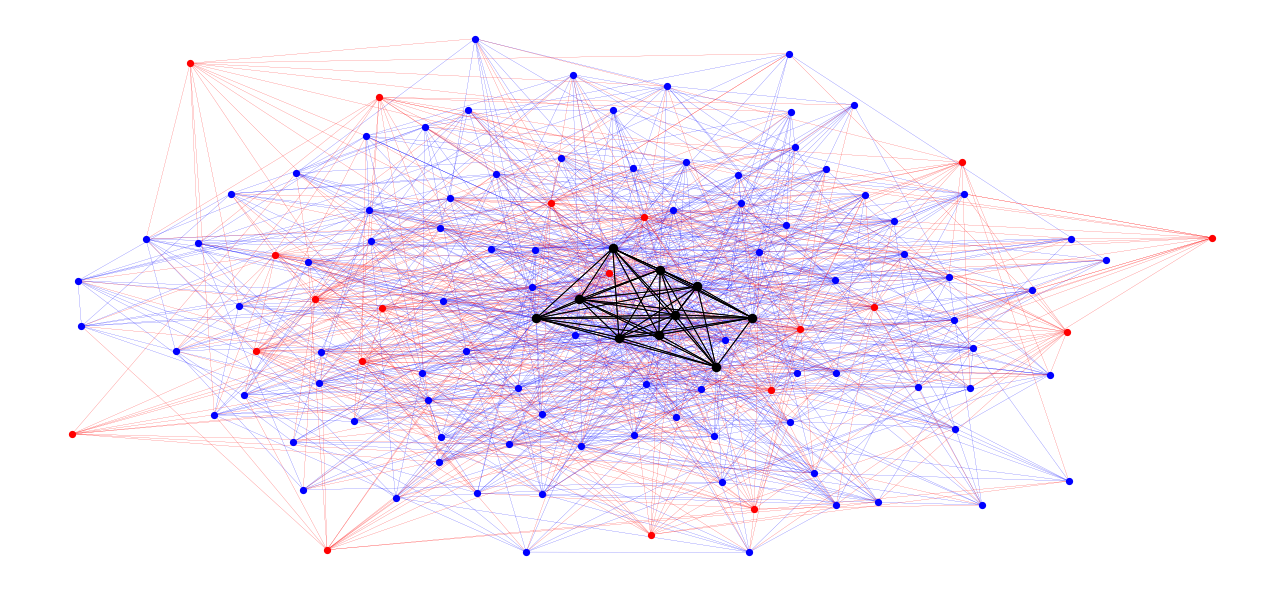} \\
{\scriptsize Setting (b), $k=50$}  \\
\includegraphics[height=.08\textheight]{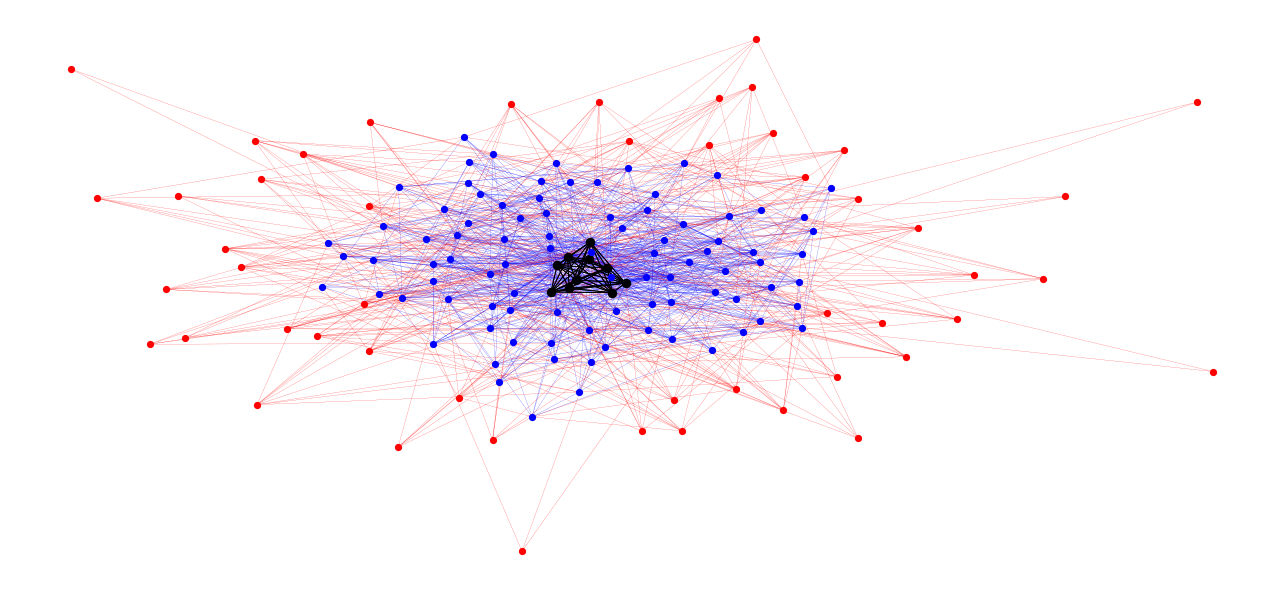} \vspace{3.35cm}
\end{center}
\end{minipage}
\vspace{-3.8cm}
\caption{Error rate of the centrality estimation methods against the latent factor dimension $k$. For the three plots on the left, we consider three settings of the sparse influence matrix ${\bm B}$; see the main text for details. The right plot depicts a realization of the graph where `{\color{red}red}'/`{\color{blue}blue}'/`black' nodes are external source/normal nodes/central nodes. As $k$ increases, setting {\sf (b)} models external influences that originate from the periphery.}\vspace{-.4cm}
\label{fig:syn_data}
\end{figure*}
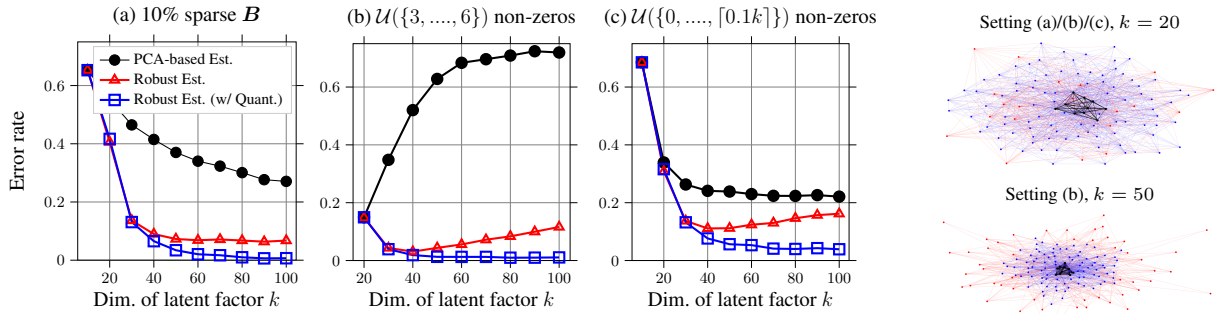

\vspace{-.1cm}
\subsection{Practical Implementation}\vspace{-.1cm}
The above analysis shows that the top singular vector of ${\bm L}$ from \eqref{eq:H=L+S} can satisfy our estimation requirement of centrality vector and we should obtain an accurate estimate of ${{\bm L}}$ from \eqref{eq:rpca}. In practice, the estimated eigen-vector obtained from $\hat{\bm L}$ is sensitive to the dimension $k$ of latent parameter ${\bm z}_i$, and the choice of  $\lambda_S, \lambda_L $ in \eqref{eq:rpca}, which may bring unstable performance to our method. To improve stability, we discuss the details for -- (a)  how to balance $\lambda_L \| \hat{\bm L} \|_\star $ and $ \lambda_S \| {\rm vec}(\hat{\bm S}) \|_1$ by choosing $\lambda_S, \lambda_L $, and (b) thresholding the sparse matrix $\hat{\bm S}$ and obtaining the centrality vector.  

Firstly, the choice of $\lambda_L$ and $\lambda_S$ in \eqref{eq:rpca} is important for maintaining the performance of our method. 
Note that the parameters trade off between low-rankness/sparseness and estimation error of ${\bm L}, {\bm S}$ in \eqref{eq:H=L+S}.
To this end, we recall from \cite[Corollary 1]{agarwal2012noisy} to set $\lambda_S = c_0 + c_1/\sqrt{k}$ and $\lambda_L = c_2$ for some $c_0,c_1,c_2 > 0$. 

Secondly, we observe from our numerical experience that with suitable $\lambda_S$ and $\lambda_L$, the estimated $\hat{\bm S}$  is relatively reliable, especially when one quantizes the $\hat{\bm S}$ obtained from \eqref{eq:rpca}. We leverage on this and \eqref{eq:H=L+S} to improve our centrality estimate. Particularly, we consider \vspace{-.1cm}
\beq \label{eq:sthres}
\widehat{\bm S}^{\sf thres} = \mathbbm{1} ( \hat{\bm S} \geq \beta ) \odot \hat{\bm S},\vspace{-.0cm}
\eeq
where $\odot$ denotes elementwise product and $\mathbbm{1}$ is an indicator function.
With $\widehat{\bm S}^{\sf thres}$, we can use $\widehat{{\bm H}{\bm B}}-\widehat{\bm S}^{\sf thres}$ as an alternative to $\hat{\bm L}$ to compute the singular vector $\widetilde{\bm v}_1$. 
We summarize the proposed centrality estimation procedure in Algorithm~\ref{alg:boosted}. Note the development of the graph filter robust centrality estimation method is inspired by \cite{wai2018blind}. Our focus here is on extracting the top eigenvector of ${\bm A}$ only, which requires higher sensitivity for the matrix estimation problem \eqref{eq:rpca} and motivated the implementation details discussed above.\vspace{-.1cm}

\section{Numerical Experiments}\vspace{-.2cm}
\label{sec:exp}
This section performs numerical experiments for proposed centrality estimation methods on synthetic and real data.\vspace{.1cm}

\noindent \textbf{Synthetic Data.} Our first goal is to verify \Cref{cor:perf} applied on graph signals that are generated from the analyzed sparse influence and GSP models. For the numerical results below, the graph $G$ of interest is generated according to a stochastic block model of two blocks with $n=100$ nodes. Let $V$ be partitioned as $V_{\sf c} = \{1,...,10\}$ and $V_{\sf p} = V \setminus V_{\sf c}$. For any $i,j \in \{1,...,n\}$, an edge is assigned independently with probability $1$ if $i,j \in V_{\sf c}$; with probability $4p$ if $i \in V_{\sf c}, j \in V_{\sf p}$; and with probability $p$ if $i,j \in V_{\sf p}$. The connectivity parameter is set as $p=0.05$. In other words, $G$ is a graph with $10$ core nodes which induce a fully connected sub-graph, while the out-of-core nodes are only sparsely connected. 
For the $n \times k$ influence matrix ${\bm B}$, we experiment with three types of sparsity patterns: {\sf (a)} 10\% of the elements in ${\bm B}$ are non-zero, selected at random, {\sf (b)} for each row of ${\bm B}$, we select ${\cal U}( \{3,...,6\})$ positions (out of $k$ possible ones) at random to be non-zero, {\sf (c)} similar to (b), but we select ${\cal U}( \{0,...,\lceil 0.1 k \rceil\})$ positions at random to be non-zero. For each selected non-zero positions, the value $B_{ij}$ is generated as a continuous r.v.~${\cal U}([0.1,1])$. Each of the above settings model a different sparse influence model, e.g., {\sf (b)} is applicable if each excitation source only influences a few nodes in the graph. 

To simulate the observed graph signals, we first generate the latent factors as ${\bm z}_i \sim {\cal N}( {\bm 0}, {\bm I}) \in \Re^k$, and consider the graph filter as ${\cal H}( {\bm A}) = ( {\bm I} - 0.1 {\bm A} )^{-1}$, and we set $m=10^5$. Finally, this generates ${\bm y}_i = {\cal H}( {\bm A} ) {\bm B} {\bm z}_i$. 
To measure the performance of centrality estimation, we let $\widehat{V}_{\sf p}$ be the set of indices that corresponds to the largest-10 indices of $| [\widetilde{v}_1]_i | $. We calculate the error rate by\vspace{-.1cm}
\beq \textstyle
{\sf Error} = \EE \big[ \frac{1}{10} | \{1,...,10\} \cap \widehat{V}_{\sf p} | \big].\vspace{-.1cm}
\eeq 
Fig.~\ref{fig:syn_data} shows the performance comparison of centrality estimation methods. In addition to the (graph filter) robust estimation methods, we compare three methods: PCA based estimation using the exact covariance ${\bm C}_y$; robust estimation based on the top singular vector of $\hat{\bm L}$ obtained by solving \eqref{eq:rpca}; and robust estimation (quantized) based on the top singular vector of $\widehat{\bm H \bm B} - \widehat{\bm S}^{\sf thres}$. Furthermore, we set $\lambda_L = 0.1$, $\lambda_S = 0.2 + \frac{2}{\sqrt{k}}$ in \eqref{eq:rpca} and $\beta = 0.1$ in \eqref{eq:sthres}.
For all three settings of ${\bm B}$, the PCA-based estimation suffers from a higher error rate than the robust methods. Meanwhile, the error rate for the boosted methods generally decreases when $k$ increases, especially when the quantization step is applied for $\hat{\bm S}$. This is reasonable as the maximum rank of $\widehat{\bm{HB}}$ is $k$, while we are interested in inferring a rank-one matrix hidden in the sum in \eqref{eq:H=L+S}. These results are consistent with our theoretical analysis.\vspace{.1cm}

\begin{table}[t]
\vspace{-.4cm}
\setlength{\tabcolsep}{2.5pt}
\centering
\scriptsize
\caption{Top-eight Estimated Central Stocks}\vspace{.2cm} \label{table:top 10}
\begin{tabular}{l l l l l l l l l}
\toprule
Method & \multicolumn{8}{c}{Top eight central stocks sorted left-to-right}\\ \midrule
PCA & NVDA & NFLX & AMZN & ADBE & PYPL & CAT & MA &GOOG \\
Robust Est. &SBUX&CSCO&COP&PEP&SLB&MCD&KO&BRK.B \\ 
Robust+Q.&GE&COP&FB&SLB&IBM&KHC&SBUX&RTN \\ 
\bottomrule
\end{tabular}\vspace{-.45cm}
\end{table}

\begin{table}[t]
\setlength{\tabcolsep}{3pt}
\centering
\scriptsize
\caption{Estimated Influence Structure}\label{table:related to B}\vspace{.2cm}
\begin{tabular}{l l l l l l l l l}\toprule
Keywords & \multicolumn{8}{c}{Top eight stocks}\\ \midrule
Trade War&WBA&BLK&PFE&FDX&MDT&F&NVDA&UNH\\ 
Sales Tax&UNH&UPS&ABBV&INTC&ORCL&LLY&DHR&BAC \\ 
Iran&KHC&LLY&GM&ORCL&UNH&HD&BLK&EXC \\ 
Oil Crisis&QCOM&WBA&F&SLB&GE&MMM&DHR&CVS\\ 
Election&GE&SBUX&CVS&MET&COST&COP&EMR&AMZN\\ \bottomrule
\end{tabular}\vspace{-.4cm}
\end{table}

\noindent \textbf{Real Data.}
We next apply our model on the daily return data from S\&P100 stocks in May 2018 to Aug 2019, consisting of $n=99$ stocks and $m=300$ samples, collected from \url{https://alphavantage.co}\footnote{The authors thank Ms.~Xuan Zhang (CUHK) for collecting the dataset.}. As described in Sec.~\ref{sec:stock}, we model each sample of daily returns as a filtered graph signal \eqref{eq:y=Hx}, \eqref{eq:x=Bz}. The latent input ${\bm z}_i$ on the relevant days is estimated from the \emph{interest level} on Google Trend (\url{https://trends.google.com}) on $k=5$ key words: `trade war', `sales tax', `Iran', `oil crisis' and `election'. We apply Algorithm~\ref{alg:boosted} to estimate ${\bm c}_{\sf eig}$ for the boosted methods.


Stocks with the top-8 estimated centrality scores are shown in Table \ref{table:top 10}, ranked from left to right. The stock influence network is likely to be clustered according to the type of industry, and therefore the stocks with high centrality should contain companies in different industries. This is the case for the estimation results of the robust methods, where stocks from food industry (SBUX), technology (FB) and energy (COP) are selected. This contrasts to the results from PCA, which returns mostly technology firms (NVDA, NFLX, etc.). 


 
We inspect the structure of the inferred influence matrix ${\bm B}$ in Table~\ref{table:related to B}. Particularly, we show the $8$ stocks with the highest magnitude in the corresponding column vectors for each keyword. These stocks should be those that are most affected by the corresponding state-of-world. 
We observe that -- `trade war'  affects stocks from the pharmaceutical industry (WBA, PEF, MDT); `sales tax' affects  technology (INTC, ORCL, etc.) stocks; `oil crisis' affects oil field (e.g., SLB) and technology (e.g., QCOM, WBA) stocks; `election' affects technology (e.g., GE, EMR, etc.) and service (e.g., CVS, SBUX, COST) stocks. 
`Iran' affects many industries like food (KHC), finance (UNH, BLK), technology (LLY, ORCL), energy (EXC) and others (GM, HD).
These results predict which stocks take the most `hit' if the state-of-world change. We remark  there maybe other factors affecting the state-of-world beyond the 5 chosen keywords.\vspace{.1cm}

\noindent \textbf{Conclusions.} 
We have studied the blind methods for centrality estimation from low-pass filtered graph signals. We establish the error bounds for two methods -- one based on PCA in folklore heuristics, and one based on a robust method. The latter is shown to be robust to unknown graph processes. We verify our findings on synthetic and real data. For future works, we will work on a complete blind setting where ${\bm z}_i$ are unknown, and the efficient algorithms for solving \eqref{eq:rpca}.


\newpage
\bibliographystyle{IEEEtran}
\bibliography{ref_list}

\begin{thebibliography}{10}
\providecommand{\url}[1]{#1}
\csname url@samestyle\endcsname
\providecommand{\newblock}{\relax}
\providecommand{\bibinfo}[2]{#2}
\providecommand{\BIBentrySTDinterwordspacing}{\spaceskip=0pt\relax}
\providecommand{\BIBentryALTinterwordstretchfactor}{4}
\providecommand{\BIBentryALTinterwordspacing}{\spaceskip=\fontdimen2\font plus
\BIBentryALTinterwordstretchfactor\fontdimen3\font minus
  \fontdimen4\font\relax}
\providecommand{\BIBforeignlanguage}[2]{{%
\expandafter\ifx\csname l@#1\endcsname\relax
\typeout{** WARNING: IEEEtran.bst: No hyphenation pattern has been}%
\typeout{** loaded for the language `#1'. Using the pattern for}%
\typeout{** the default language instead.}%
\else
\language=\csname l@#1\endcsname
\fi
#2}}
\providecommand{\BIBdecl}{\relax}
\BIBdecl

\bibitem{kolaczyk2014statistical}
E.~D. Kolaczyk and G.~Cs{\'a}rdi, \emph{Statistical analysis of network data
  with R}.\hskip 1em plus 0.5em minus 0.4em\relax Springer, 2014, vol.~65.

\bibitem{friedman2008sparse}
J.~Friedman, T.~Hastie, and R.~Tibshirani, ``Sparse inverse covariance
  estimation with the graphical lasso,'' \emph{Biostatistics}, vol.~9, no.~3,
  pp. 432--441, 2008.

\bibitem{timme2007revealing}
M.~Timme, ``Revealing network connectivity from response dynamics,''
  \emph{Physical review letters}, vol.~98, no.~22, p. 224101, 2007.

\bibitem{wai2016active}
H.-T. Wai, A.~Scaglione, and A.~Leshem, ``Active sensing of social networks,''
  \emph{IEEE Transactions on Signal and Information Processing over Networks},
  vol.~2, no.~3, pp. 406--419, 2016.

\bibitem{ravazzi2017learning}
C.~Ravazzi, R.~Tempo, and F.~Dabbene, ``Learning influence structure in sparse
  social networks,'' \emph{IEEE Transactions on Control of Network Systems},
  vol.~5, no.~4, pp. 1976--1986, 2017.

\bibitem{wai2019joint}
H.-T. Wai, A.~Scaglione, B.~Barzel, and A.~Leshem, ``Joint network topology and
  dynamics recovery from perturbed stationary points,'' \emph{IEEE Transactions
  on Signal Processing}, vol.~67, no.~17, pp. 4582--4596, 2019.

\bibitem{giannakis2018topology}
G.~B. Giannakis, Y.~Shen, and G.~V. Karanikolas, ``Topology identification and
  learning over graphs: Accounting for nonlinearities and dynamics,''
  \emph{Proceedings of the IEEE}, vol. 106, no.~5, pp. 787--807, 2018.

\bibitem{matta2018consistent}
V.~Matta and A.~H. Sayed, ``Consistent tomography under partial observations
  over adaptive networks,'' \emph{IEEE Transactions on Information Theory},
  vol.~65, no.~1, pp. 622--646, 2018.

\bibitem{sandryhaila2013discrete}
A.~Sandryhaila and J.~M. Moura, ``Discrete signal processing on graphs,''
  \emph{IEEE Trans. Signal Process.}, vol.~61, no.~7, pp. 1644--1656, 2013.

\bibitem{shuman2013emerging}
D.~Shuman, S.~Narang, P.~Frossard, A.~Ortega, and P.~Vandergheynst, ``The
  emerging field of signal processing on graphs: Extending high-dimensional
  data analysis to networks and other irregular domains,'' \emph{IEEE Signal
  Process. Mag.}, vol.~3, no.~30, pp. 83--98, 2013.

\bibitem{marques2015sampling}
A.~G. Marques, S.~Segarra, G.~Leus, and A.~Ribeiro, ``Sampling of graph signals
  with successive local aggregations,'' \emph{IEEE Transactions on Signal
  Processing}, vol.~64, no.~7, pp. 1832--1843, 2015.

\bibitem{anis2016efficient}
A.~Anis, A.~Gadde, and A.~Ortega, ``Efficient sampling set selection for
  bandlimited graph signals using graph spectral proxies,'' \emph{IEEE
  Transactions on Signal Processing}, vol.~64, no.~14, pp. 3775--3789, 2016.

\bibitem{sandryhaila2014big}
A.~Sandryhaila and J.~M. Moura, ``Big data analysis with signal processing on
  graphs,'' \emph{IEEE Signal Processing Magazine}, vol.~31, no.~5, pp. 80--90,
  2014.

\bibitem{dong2016learning}
X.~Dong, D.~Thanou, P.~Frossard, and P.~Vandergheynst, ``Learning laplacian
  matrix in smooth graph signal representations,'' \emph{IEEE Transactions on
  Signal Processing}, vol.~64, no.~23, pp. 6160--6173, 2016.

\bibitem{segarra2017network}
S.~Segarra, A.~G. Marques, G.~Mateos, and A.~Ribeiro, ``Network topology
  inference from spectral templates,'' \emph{IEEE Transactions on Signal and
  Information Processing over Networks}, vol.~3, no.~3, pp. 467--483, 2017.

\bibitem{egilmez2017graph}
H.~E. Egilmez, E.~Pavez, and A.~Ortega, ``Graph learning from data under
  laplacian and structural constraints,'' \emph{IEEE Journal of Selected Topics
  in Signal Processing}, vol.~11, no.~6, pp. 825--841, 2017.

\bibitem{dong2019learning}
X.~Dong, D.~Thanou, M.~Rabbat, and P.~Frossard, ``Learning graphs from data: A
  signal representation perspective,'' \emph{IEEE Signal Process. Mag.},
  vol.~36, no.~3, pp. 44--63, 2019.

\bibitem{mateos2019connecting}
G.~Mateos, S.~Segarra, A.~G. Marques, and A.~Ribeiro, ``Connecting the dots:
  Identifying network structure via graph signal processing,'' \emph{IEEE
  Signal Process. Mag.}, vol.~36, no.~3, pp. 16--43, 2019.

\bibitem{wai2018blind}
H.-T. Wai, S.~Segarra, A.~E. Ozdaglar, A.~Scaglione, and A.~Jadbabaie, ``Blind
  community detection from low-rank excitations of a graph filter,''
  \emph{arXiv:1809.01485v2}, 2018.

\bibitem{hoffmann2018community}
T.~Hoffmann, L.~Peel, R.~Lambiotte, and N.~S. Jones, ``Community detection in
  networks with unobserved edges,'' \emph{arXiv preprint arXiv:1808.06079},
  2018.

\bibitem{schaub2019blind}
M.~T. Schaub, S.~Segarra, and J.~N. Tsitsiklis, ``Blind identification of
  stochastic block models from dynamical observations,''
  \emph{arXiv:1905.09107}, 2019.

\bibitem{yuan2019learning}
Y.~Yuan, H.~H. Yang, and T.~Q. Quek, ``Learning overlapping community-based
  networks,'' \emph{IEEE Trans. Signal Inf. Process. Netw.}, 2019.

\bibitem{horn2012matrix}
R.~A. Horn and C.~R. Johnson, \emph{Matrix analysis}.\hskip 1em plus 0.5em
  minus 0.4em\relax Cambridge university press, 2012.

\bibitem{vigna2016spectral}
S.~Vigna, ``Spectral ranking,'' \emph{Network Science}, vol.~4, no.~4, pp.
  433--445, 2016.

\bibitem{agarwal2012noisy}
A.~Agarwal, S.~Negahban, M.~J. Wainwright \emph{et~al.}, ``Noisy matrix
  decomposition via convex relaxation: Optimal rates in high dimensions,''
  \emph{The Annals of Statistics}, vol.~40, no.~2, pp. 1171--1197, 2012.

\end{thebibliography}

\end{document}